# A symbolic-arithmetic for teaching double-black node removal in red-black trees

Kennedy E. Ehimwenma[0000−0002−7616−9342], Junfeng Wang, Ze Zheng, and Hongyu Zhou

Department of Computer Science, College of Science and Technology,
Wenzhou-Kean University,
88 Daxue Rd, Ouhai, Wenzhou, Zhejiang Province, China 325060 {kehimwen, wangjunf, zhenze, zhouho}@kean.edu

**Abstract.** A red-black (RB) tree is a data structure with red and black nodes coloration. The red and black color of nodes make up the prin-cipal component for balancing a RB tree. A balanced tree has an equal number of black nodes on any simple path. But when a black leaf node is deleted, a double-black (DB) node is formed, thus, causing a reduction in black heights and the tree becomes unbalanced. Rebalancing a RB tree with a DB node is a fairly complex process. Teaching and learning the removal of DB nodes is also challenging. This paper introduces a simplified novel method which is a symbolic-algebraic arithmetic proce-dure for the removal of DB nodes and the rebalancing of black heights in RB trees. This simplified approach has enhanced student learning of the DB node removal in RB trees. Feedback from students showed the learn-ability, workability and acceptance of the symbolic-algebraic method in balancing RB trees after a delete operation.

## 1    Introduction

Red-black (RB) trees are binary trees and by extension a binary search tree [7, 14, 19]. In a RB tree, the color of a node is either red or black. Any node, whether red or black, can be deleted due to data update or modification in the data structure. Fair enough, the deletion of a red node and tree rebalancing is straightforward as a deleted red node is replaced by a black node: an operation that we best explained as ***Red*** + ***Black*** = ***Black***. But the deletion of a black node or the replacement node (that was moved from its position to the place of a deleted ***Red node***) is where lies the complexity of recoloring and rebalancing of the RB tree. This paper considers a new approach for handling the complexity of recoloring and rebalancing in RB trees. Our approach is a step-by-step simulation of algebraic operation which involves the use of symbols: **R** (for red), **B** (for black), and **DB** (for double-black). Addition and subtraction of these symbolic colors are applied as the case may be in the balancing of the tree.

In classroom teaching and learning, the complexity of the red-black tree recoloration and rebalancing poses a challenge to both tutors and students alike [23]. This is because: *1) the deletion of a black node reduces the number of black nodes along any simple path from the root node to any descendant leafnode; then, 2) a double-black node is created which replaces the deleted black node: this also explained as **B + NULL-LEAF = DB**; and 3) the double-black node is needed to be removed from the red-black tree*. The removal of this double-black node is where lies the complexity.

The deletion of a black node and its replacement by a double-black, symbol-ically as, **B + NULL-LEAF = DB** reduces black heights in the tree. Black height refers to the number of black nodes on any path from the root node to any descendant leafnode. When black height is reduced and black nodes becomes unequal along any path in comparison to other paths, the RB tree becomes un-balanced.

That is, the tree requires a rebalance such that the DB node is removed from any path in the tree to have an equal number of black nodes. Double-black nodes have no place in RB trees. Yet when a black node is deleted, a DB node is formed. The conventional algorithm for removing DB nodes poses a strong challenge to students. While there are several scholarly works on node recoloring and balancing of the RB tree; not many works have addressed the problem of DB removal. This is the problem that this paper has addressed using symbolic arithmetic computation to remove the DB node and to present a new teaching approach to ease DB removal and tree rebalancing, subsequently.

## 1.1 Contributions

If there is an occurrence of a DB node as a result of a deleted node in a RB tree; how is the DB node removed? How are the nodes reassigned their colors after the removal of a DB node? From the foregoing the contribution of this paper are, namely, 1) to present a simplified symbolic-arithmetic procedure for teaching and learning DB node removal in computer science curriculum, 2) to demonstrate a symbolic-arithmetic approach for the recoloring and balancing of the RB tree, and 3) to project the supporting algorithm for the symbolic-algebraic arithmetic procedure for DB node removal and tree rebalancing. This article continues with section 2 as related works on the RB tree data structure; section 3 presents our symbolic-arithmetic methodology that is further simplified by an algorithm. Section 4 discusses our symbolic arithmetic operation using illustrative RB tree diagrams. In addition is students' feedback w.r.t. our symbolic-algebra method in comparison to the conventional RB tree DB removal algorithm. Section 5 is conclusions and further works.

## 1.2 Statement of the problem

The deletion of a black node in a RB tree causes a reduction in black heights, tree rotation and restructuring, as well as recoloring of nodes. This is a fairly complex and challenging algorithm to learn or teach in the process of balancing a RB tree. The statement of the problem thus states: *There is a mathematically-based system of approach that can be used to teach and ease the learning of the conventional algorithm of double-black removal, and the subsequent recoloring of nodes and balancing of the RB tree.*

## 2    Related works

A RB tree is one of several binary search trees. In data structures, every binary tree has their peculiar properties that defines, namely; its structure, height and balancing of the tree. One of the attributes of the RB tree is the "color" field in which every node in the tree is assigned a "color" which is either red or black [7, 11, 14]

### 2.1    Properties of red black trees

a) Each node is either red or black.
b) The root node is black.
c) Each leaf node is black.
d) The children of a red node are black.
e) For each node, all simple paths from the node to descendent leaves contain      the same number of nodes.
f) Two consecutive nodes cannot be both red.
g) A red-black tree is a binary search tree.

At the point when all of these requirements are met, a red-black tree is created. Furthermore, if the insert or delete operation (or method) is called for a given node, the structure of the tree and nodes' color also changes to reflect the requirements which helps the tree to rebalance itself [3]. Based on node coloration; the subtrees (or children) of a black node can; *i) both be blacks, ii) red and black, or iii) both reds* (figure 1). But for a red node, its two children must always be black [23]. This is because a red node must have a black parent as well as a black child node. Otherwise, the property that states *"no two nodes that are connected side-by-side can be red"* would be violated.

After an operational manipulation such as insertion, update or delete; data structures like the AVL and RB trees must be balanced [8]. A RB tree is bal-anced if a simple path from the root node to every descendant leafnode has equal number of black nodes (figure 1). On the other hand; it is unbalanced if there are unequal number of black nodes on every simple path from the root to the leafnodes. The RB tree is a famous data structure for the storage and manage-ment of data. In the non-teaching fields, several studies e.g. [10, 13, 18, 25] have been conducted on the impact analysis of RB trees on memory management and performance. Also [17] conducted a research on the application of RB trees in wireless sensor networks; and the use of RB trees for optimising costs in network trees with a prescribed algorithm [9].

The RB tree is an important topic in data structures in the study of computer science that students can find very challenging and difficult to learn [9, 15, 22].

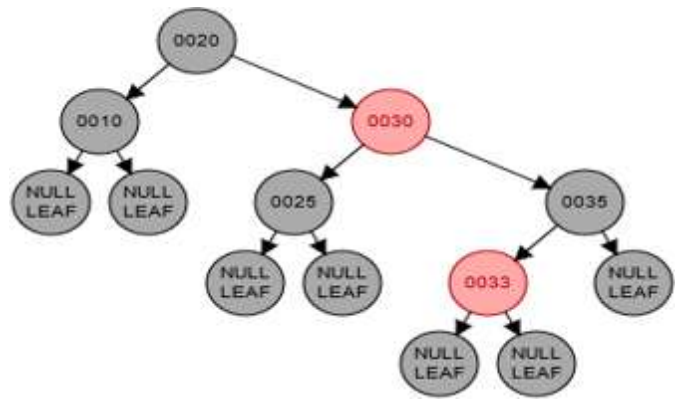

**Fig. 1.** A balanced red-black tree showing null leaves. In a Red-black visualizer [4].

As an abstract data type, that is aimed at understanding computer storage and software engineering; the operations performed on a RB are insertion, rotation, deletion, and recoloring of nodes. In looking out for new approaches to teaching RB trees, Wu et al. [22] proposed a customised platform for learning data struc-tures after stating series of steps to learn data structures with-ease but with no mention of topical areas in the subject. To improve the efficiency of teaching data structures, Seidametova [21] conducted experiments using different student groups e.g. Group1 vs. Group2 to compare the effectiveness of teaching strategy using visualization tool vs. flipped-classroom in the areas of Hashing and Trees (BST, RBT, AVL). The scores obtained were used to determine what group learned the most. King [12] reported how their university data structures & algorithm was redesigned to reflect experimental analysis. The revised curricu-lum and practice, thereafter, enabled students to include experimental analysis in their studies to connect computer science theory with software engineering practice. The report methodology was concluded with the use of Likert Rat-ing Scale for the collection of metrics. The report of Nipkow [16] presented a list of the most of the conventional topics in data structures & algorithms; and discussed that algorithms are logics. Thus, data structures course needs to be supported with critical computational thinking and formal proofs and logic. In [23] a top-down insertion method was projected to address the problem of single and double rotation through a granularity approach in order to balance the RB after insertion. The granularity approach prescribed a step-by-step selection of rules for students to follow and to balance the RB tree. Several approaches for addressing the teaching and/or learning of RB have been presented but most of which are in the rotation and recoloring of nodes e.g. [2, 23] which did not make the problem any less.

Whereas, the insertion operation of a new node in a RB tree is the simpler of two hard problems: With the harder being the deletion of a black node or the replacement node [6] and the subsequent rebalancing of the tree. In order to address the removal of DB node, Zegour [24] used formalized statements in the description of nodes' color, recoloring and tree balancing without simplifying the challenges in students' learning of the DB removal algorithm. In [6] a parity-seeking delete algorithm was introduced with the goal similar to the aim of our paper: to introduce a pedagogically sound and easy way to understand the algorithm for RB tree deletion algorithm. The rationale which is to balance either by repairing a defective subtree also left a bit of complexities in understanding. The work of Sedgewick [20] presented 2-3 variants of the RB tree, called the Left-Leaning Red-Black trees (LLRB), and proposed concise number of deletion algorithms. However, the deletion algorithm of the LLRB is still complex and not suitable for education. Based on these complexities that is left with the deletion algorithm of the RB tree, the aim of this paper is not about insertion of node as several studies have been carried out on node insertion – which is pretty straight forward. But on DB node removal using basic arithmetic operations after a delete operation. On the deletion of a black node; the conventional DB removal algorithm is not deterministic in approach e.g. see relevant chapters in [7, 14] and visualization tools [4]. Hence, the quest for a new mathematically-based algebraic model of operation.

## 3    Methodology

This paper presents a simple arithmetic process for removing the DB node and rebalancing of the RB tree. Our strategy is technically a simple addition or subtraction of the red *R* or black *B* color to/from an existing color of a given node; like (–) × (–) = (+) in an algebraic operation. Symbolically, we subtract a single black, *–B*, color from the two siblings of a node – of which one is a DB node – and then add a single black, *+B*, to the parent of the two siblings. Such that we have, *–B*(*leftChild*), *–B*(*rightChild*) and *+B*(*parent*); (figure 2).

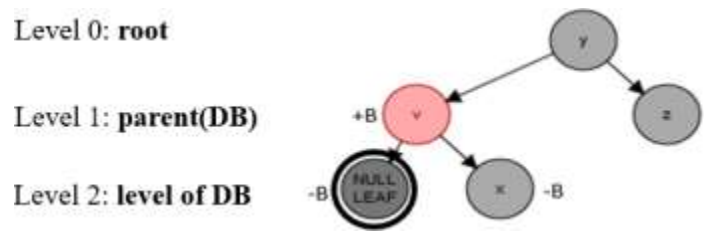

**Fig. 2.** Symbolic *subtraction* of the *color black* (represented as *–B*) from two siblings and the *addition* of color *+B* to their *parent v*.

### 3.1 The algebraic arithmetic of double-black node removal and tree balancing

**The symbolic-algebraic arithmetic rules.** Our symbolic-arithmetic opera-tion in the process of removal of a DB node in a RB tree after a delete operation on a black node so as to rebalance the black-height of the tree are given as:

| | |
|---|---|
| black + black = **double black** | *Eq.* (1) |
| **double black** – black = black | *Eq.* (2) |
| black – black = red | *Eq.* (3) |
| red + black = black | *Eq.* (4) |
| red + **double black** = black | *Eq.* (5) |
| red – black = black | *Eq.* (6) |

The given operations imply the simple addition of red or black color to an existing node's color in order to obtain the resultant node color in the process of DB removal and black heights rebalancing of the tree. It should be noted that the color ***operands*** *in the formulas above cannot be moved to the other side of the equal sign. Otherwise, conflicts may occur.*

**The change factor** is the color added or subtracted to/from the original color. For example, if a node is initially black and an extra black color is added to it, we have as stated in *Eq.* (1):

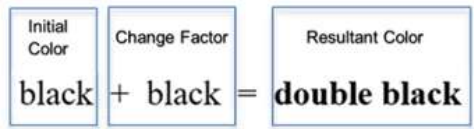

**Fig. 3.** An example of the change factor in equation (1).

**The transmission rules of the change factor:**

a) The path, or order, of the recoloring process travels from the DB node to its adjacent nodes; either along or against the directed edges to the deleted node.
b) Start the symbolic-algebraic rule application at the level of the DB. c) The symbolic-algebraic rule application is always upward towards the root. d) The traversal stops when either of these condition is reached: A DB node is    generated at the root node, or the tree is already balanced.

### 3.2   Delete operation and double-black node

In a RB tree, every simple path from the root node to any descendant leafnode must have equal number of black nodes. This means that the deletion of a red leafnode does not affect the number of black nodes along any simple path in a tree. There is however a problem, if an external black leafnode is deleted, or if an internal red node is deleted and then replaced by an external black leafnode. To correct a DB node occurrence, Besa and Eterovic [1] described it as the removal of a black unit from say node *A* and passing it to a different node *B* and if node *B* is black then it will have an extra black. A tree that has a DB node is not a RB tree [5]. Thus, to remove DB nodes, Germane and Might [5] introduced the operation of DB node rotation.

## 4   Discussion

In this section we present our technical but simple algebraic rules for the removal of DB nodes that subsequently leads to the balancing of any RB tree as we recur up the tree from children to parent. A parent node has two sub-trees, namely, the left and right subtrees.

### 4.1   Recursive subtraction and addition of black color

Irrespective of whether a DB node is a left subtree or right subtree, our algebraic algorithm *subtracts 1 black color* i.e. *–B* from both subtrees; and *add 1 black color* *+B* to the parent node. Bottom-up the tree, and along the path-traversal of a given DB node towards the root of the tree; this process recursively continues if there is a re-occurrence of another DB node (by virtue of *+B* addition to a parent black node) unless the node is a root.

**Deletion of a red leafnode.** As shown in figure 4 (a & b), the deletion of the leaf nodes 15 and 33 respectively is a ***simple case***. This deletion operation does not in any way affect black heights in the tree nor has it made the tree unbalanced. However, as shown in the following section, a delete operation on an internal red node or on any black node causes an imbalance in RB trees. For our illustrations, we have used the following parameters to represent the nodes in any given subtree and their parent. Thus,
   *P* for the parent of double-black node
   *U* for the double-black, and
   *S* as sibling of the double-black node.
   For discussion purposes, we shall use the notations, such as, *B(P)* for a black parent node, *DB(U)* for the double-black node, and *B(S)* or *R(S)* for a black or red sibling of the DB node, respectively. In addition, the parameters *P*, *U* and *S* shall be substituted, where necessary, with their respective node values. It is important to note that DB nodes are NOT only produced from the deletion of black leaf nodes alone but are also formed in the position of a black leaf node that is replacing an internal red node that has been deleted (as explained in section "Deletion of an internal red node", figure 7).

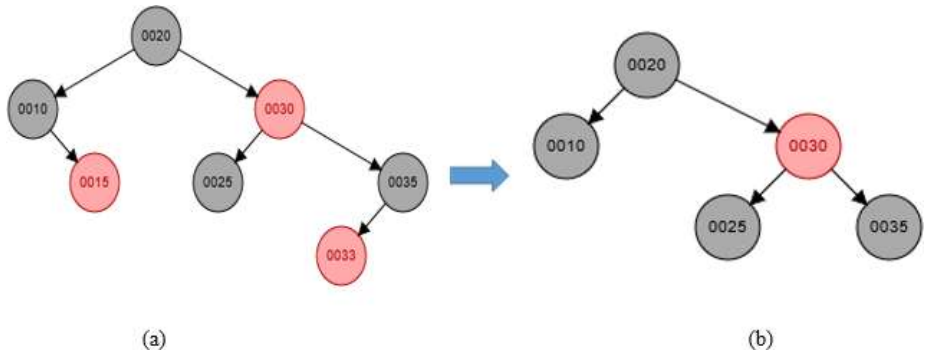

**Fig. 4.** In (a) the deletion of the red leafnodes of 15 and 33, respectively; results in figure 4(b).

**Deletion of a black leafnode.** Since the node 30 is a root node, and since every root node must be color black, then node 30 is black. Otherwise, we recur up the tree by re-applying the symbolic expression *–B* to both the **DB** node and its sibling *s* respectively, and *+B* to their parent *p* node as shown in figure 2.

*Illustration 1:*

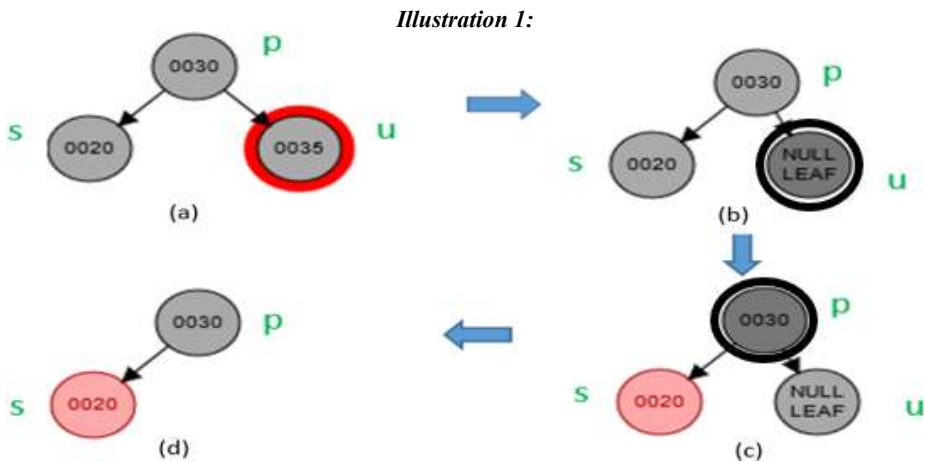

**Fig. 5.** (a, b, c, d): Deletion of black node 35 which left a double-black and black height rebalancing.

**Table 1.** Steps of balancing and color assignment for case "Deletion of a black leafnode" illustration 1.

| Steps | Description | Operated node | Initial color | Operation | Eq. used | Change factor | Final color | Tree balanced or not |
|---|---|---|---|---|---|---|---|---|
| 1 | To remove double black on null leaf | DB Node | **DB** | **DB** – black = black | (2) | – black | Black NULL | NO |
| 2 | Inverse the change factor and apply to Node 30 to balance | 30 | black | black(30) + black = **DB** | (1) | + black | **DB(30)** | NO |
| 3 | Inverse the change factor and apply to Node 20 to balance | 20 | black | black(20) – black = red | (3) | – black | red(20) | YES |

*Illustration 2:*

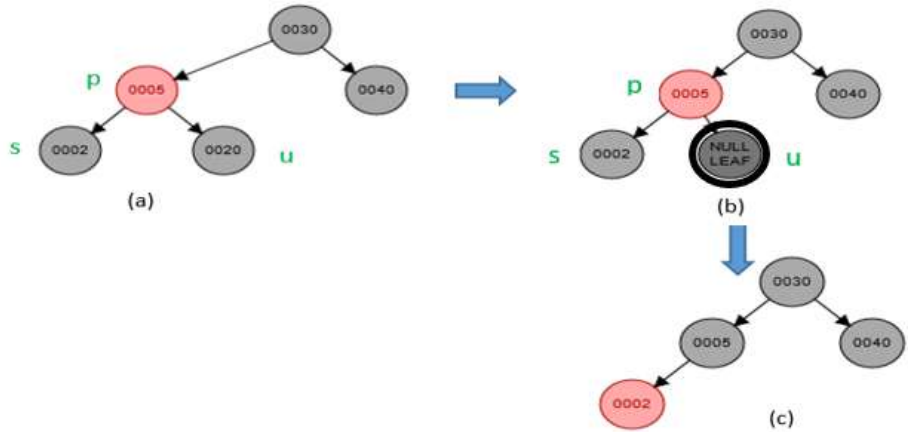

**Fig. 6.** (a, b, c): Deletion of black node 20 which left a double-black and black-height rebalancing.

**Table 2.** Steps of balancing and color assignment for case "Deletion of a black leafnode" illustration 2.

| Steps | Description | Operated node | Initial color | Operation | Eq. used | Change factor | Final color | Tree balanced or not |
|---|---|---|---|---|---|---|---|---|
| 1 | To remove double black null leaf | DB Node | **DB** | **DB** – black = black | (2) | – black | black NULL | NO |
| 2 | Inverse the change factor and apply to Node 5 to balance | 5 | red | red(5) + black = black | (4) | + black | black(5) | NO |
| 3 | Inverse the change factor and apply to Node 2 to balance | 2 | black | black(2) – black = red | (3) | – black | red(2) | YES |

**Deletion of an internal red node.** If the internal red node 25 is deleted, the value of this deleted position is replaced by the value of left child 15 which is still keeping the red color of the deleted node; but causing a DB node in the position of the replacement node. At this point in time, structurally, this is equivalent to deleting a black left child 15 and then producing a DB node. In both scenarios, our symbolic-algebraic arithmetic holds: *–B(U )*, *–B(S)* and *+B(P)*. Therefore, the problem turns out to be the case in section "Deletion of a black leafnode" Illustration 2 above.

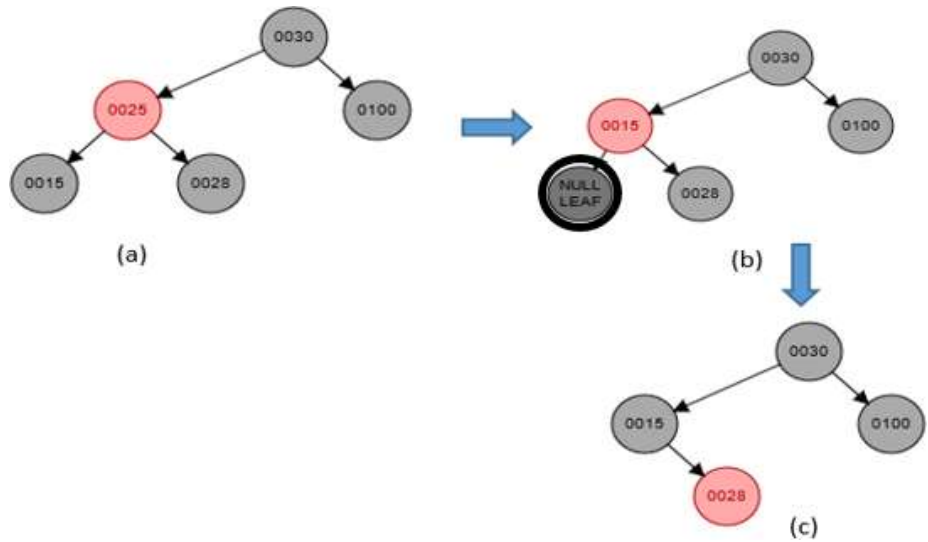

**Fig. 7.** (a, b, c): Deletion of red node 25 and replacement by node 15 left a double-black in the position of node 15 and subsequent black-height rebalancing.

**Table 3.** Steps of Balancing and Color Assignment for case "Deletion of an internal red node".

| Steps | Description | Operated node | Initial color | Operation | Eq. used | Change factor | Final color | Tree balanced or not |
|---|---|---|---|---|---|---|---|---|
| 1 | To remove double black null leaf | DB Node | **DB** | **DB** – black = black | (2) | – black | black NULL | NO |
| 2 | Inverse the change factor and apply to Node 15 to balance | 15 | red | red(5) + black = black | (4) | + black | black(15) | NO |
| 3 | Inverse the change factor and apply to Node 28 to balance | 28 | black | black(28) – black = red | (3) | – black | red(28) | YES |

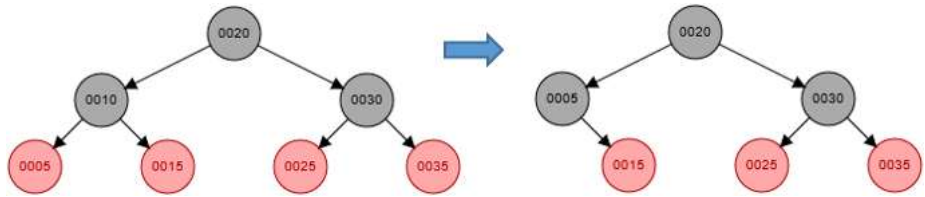

**Fig. 8.** Deletion of an internal black node 10, and replacement by node 5.

**Deletion of an internal black node.** According to the binary search tree deletion rule, the deleted node 10 is replaced by its rightmost child of the left subtree – in this case it is the left child 5 – which is the only subtree. As shown in figure 8, the deletion of the internal black node 10 is structurally the same as deleting a red leaf node on the ground that there is no DB formation, which is a *simple case* rule ***B + R = B*** (*Eq.* (4)).

Without a DB formation, our symbolic-algebraic color rule ***B + R = B*** only applies to the deleted node 10 in this case, and the replacement node 5 becomes NULL LEAF. Thus, the problem turns out to be the situation in section"Deletion of a red leafnode".

**Deletion of a root node.** By the deletion rule of the binary search tree, as applied in figure 8, the deleted node 20 is replaced by its rightmost (largest) node of the left subtree node 15 with its color unchanged. The reason is not just only because node 20 is the root but also by applying our color rule ***B + R = B***. As shown in figure 9, the deletion of the root node 20 is structurally the same as deleting a red leaf node, which is again a *simple case* rule ***B + R = B***. Therefore, the problem again turns to the situation in section "Deletion of a red leafnode".

*Illustration 1:*

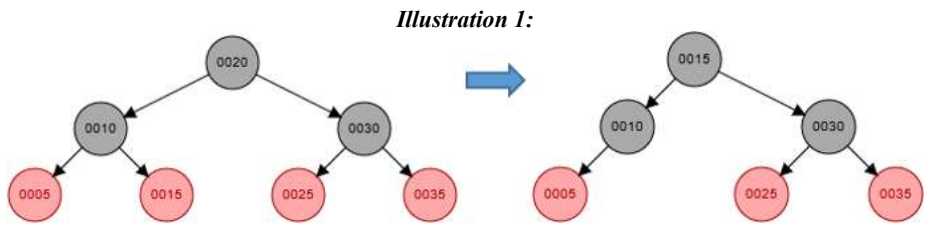

**Fig. 9.** Deletion of the root node 20; and replacement by red node 15 which is the rightmost of the root's left subtree.

When the root node 15 is deleted; the value of this position is replaced by its largest left child 10, and the color remains unchanged. Although 15 is a

***Illustration 2:***

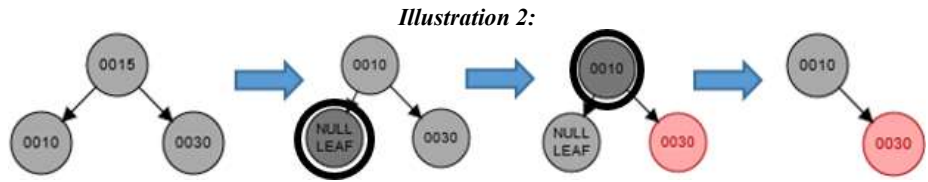

**Fig. 10.** Deletion of the root node 15; and replacement by black node 10 which is the root's left subtree.

root node, which is always black, yet we can apply the our symbolic-algebraic rule. Firstly, ***B(root) + B = DB(root)*** after the delete operation on node 15 (*Eq.* (1)). Secondly, applying (*Eq.* (2)) we can arrive at a black root node from ***DB(root) – B = B***. Here we prove the application of the symbolic-algebraic rules as given in table 4 in the process of balancing the tree. At this point, structurally, it is equivalent to deleting a black left child to produce a double-black NULL LEAF.

To apply our rules, we call the ***DB(U) – B = B***; ***B(30) – B = R***; and ***B(10) + B = DB(10)*** respectively. Of course as a root node; ***DB(10)*** becomes black. The problem is similar to the situation in section "Deletion of a black leafnode" Illustration 2. Except that here in table 4 (Steps 3 & 4), we are comparing the use of two steps that involves the color operations ***+R*** and its inverse ***–R*** as against one step of figure 5 (Step 3) which is ***B – B = R***. So, our take here is to de-emphasize the use of any additional steps as this would incur *overhead*. Table 5 reduces this overhead and it is more efficient to apply ***B – B = R*** which is a lesser step.

**Table 4.** Steps of balancing and color assignment for case "Deletion of a root node" illustration 2.

| Steps | Description | Operated node | Initial color | Operation | Eq. used | Change factor | Final color | Tree balanced or not |
|---|---|---|---|---|---|---|---|---|
| 1 | To remove double black null leaf | DB Node | **DB** | **DB** – black = black | (2) | – black | black NULL | NO |
| 2 | Inverse the change factor and apply to Node 10 to balance | 10 | black | black(10) + black = **DB** | (1) | + black | **DB(10)** | NO |
| 3 | Remove double black on node 10 | 10 | **DB** | **DB** + red = black | (5) | + red | black(10) | NO |
| 4 | Inverse the change factor and apply to Node 30 to balance | 30 | black | black(30) – red = red | (6) | – red | red(30) | YES |

**Table 5.** A more efficient steps to balancing black height and color assignment for case "Deletion of a root node" illustration 2.

| Steps | Description | Operated node | Initial color | Operation | Eq. used | Change factor | Final color | Tree balanced or not |
|---|---|---|---|---|---|---|---|---|
| 1 | To remove double black null leaf | DB Node | **DB** | **DB** – black = black | (2) | – black | black NULL | NO |
| 2 | Inverse the change factor and apply to Node 10 to balance | 10 | black | black(10) + black = **DB** | (1) | + black | **DB(root)** | NO |
| 3 | Inverse the change factor and apply to Node 30 to balance | 30 | black | black(30) – black = red | (6) | – red | red(30) | YES |

### Algorithm of double-black removal by symbolic-algebraic operation

```
 1: function DoubleBlack Removal(T, x):
 2:      if root = DB then
 3:          DB(root) – B ⇒ B(root)
 4:      end if
 5:      if x is deleted and x is a root node then
 6:          color(x) = B
 7:      end if
 8:      if color(x) = R then
 9:          if R(x) is deleted then
10:              R(x) + B ⇒ B(x)
11:          end if
12:      end if
13:      if color(x) = B then
14:          if B(x) is deleted then
15:              B(x) + B ⇒ DB(x)
16:          end if
17:      end if
18:      /** removal of DB node */
19:      while T is not balanced do
20:          if DB(x) has a sibling then
21:              DB(x) – B ⇒ B(x) && (B(s(x)) – B ⇒ R(s(x)) || R(s(x)) – B ⇒ B(s(x)))   if
22:          parent(x) = R then
23:              R(p(x)) + B ⇒ B(p(x))
24:          end if
25:          if parent(x) = B then
26:              B(p(x)) + B ⇒ DB(p(x))
27:          end if
28:          end if
29:          /** recur up the tree */
30:      end while
31:      Output Tree
32: end function
```

### 4.2 Algorithm

This is the algorithmic representation of symbolic-algebraic operation. Using first-order logic (FOL), the FOL notations are depicted as follows: $B(s(x))$ and $R(s(x))$ as the black and red sibling of the deleted node; and $B(p(x))$ and $R(p(x))$ as the black and red parent of the deleted node, respectively; while $DB(x)$ is the double-black node. As we recur up the tree, a new DB is created, and the new DB gets a new sibling and parent.

### 4.3 Feedback from the symbolic algebraic teaching approach

For over three academic years this symbolic-algebraic (SA) method have been devised to assist students with the learning of double-black (DB) removal and RB tree rebalancing. In and out of class; two fundamental questions that were put before students are stated in the following subsection in Question 1 and 2. Below each of the questions are the tables 6 and 7, respectively, which depicts students' feedback as well as our analysis of these feedbacks. In the feedback analysis we have used the acronym, SA = symbolic-algebra, and TA = traditional (conventional) algorithm.

**Question 1:** *What is your opinion on this new symbolic method in the understanding of the removal of DB and balancing of the tree?*

Table 6. Feedback on the understanding of the symbolic-algebraic method.

| Student feedback | Analysis |
|---|---|
| *For me, the new [symbolic] algebra (SA) algorithm is easier to understand compared to conventional approach, with less classifica-tions and more symbolic visual aids. But it still requires practicing to get the hang of it.* | * SA approach is simpler. <br> * SA is visual and animated in process. <br> * SA has less classification. <br> * SA is needed to be learned too. |
| *Symbolic algebraic application to tree bal-ance after rotations and node coloration has a fixed pattern to follow which is lacking in the traditional algorithm.* | * SA provides a fixed length of steps. |
| *The symbols method provides better under-standing for node coloration. It explains what node needs a color change, why it needs to change and how it will change.* | * SA gives clearer understanding. <br> * SA gives the step and the exact node to color. |

**Question 2:** *What are the perceived differences to other approaches in literature and the original DB removal algorithm?*

**Table 7.** Feedback on the conventional RB tree algorithm vs. the symbolic-algebraic method.

| Student feedback | Analysis |
|---|---|
| *This [symbolic approach] is definitely sim-pler for me because it simplifies the process inherent in a traditional algorithm (TA). Both the traditional red-black tree algorithm and the new symbolic approach algorithm involves remembering several conditions for deletion and rotation before carrying out cor-responding operations. In this case [symbolic approach], less is definitely better, because it's not as complicated as it used to be.* | * SA is definitely simpler.<br>* Both the TA and new SA have cer-tain conditions to learn (*w.r.t.* rota-tion); but the SA is lesser with infor-mation.<br>* SA method is not complicated. |
| *Traditional algorithm is somehow compli-cated. Not knowing where to stop color ex-change between nodes sometime. The sym-bolic operation is a much simplified method of tree balancing after a node removal. The symbolic method uses a step-by-step mathe-matical approach that is certain about when to end color change.* | * TA algorithm is complicated.<br>* TA is difficult to predict where color change will end.<br>* SA algorithm is much simpler.<br>* SA is step-by-step. |
| *The symbols approach takes away the com-plications involved in the heavy informa-tion that surrounds the traditional algo-rithm. This is because there are few informa-tion to process as we fix the colors of nodes with the symbols method. The steps are easy to follow.* | * SA is not complicated.<br>* SA has less information *w.r.t.* its op-eration. |
| *The deletion operations of traditional red-black [algorithm] are badly explained in many resources I read including my textbook. These operations are not obvious to under-stand and confuse many students when they first see it. The new symbolic method pro-vides steps to memorize the operations and makes removal of DB easier to learn.* | * TA is not clear to understand.<br>* SA is a step-by-step operation.<br>* SA makes DB removal easier to memorize and learn. |

### 4.4   Research findings

Therefore, from our methodology, illustrations, and application of the symbolic-algebra operation; we affirm to the **statement of problem** in section 1.2 that *there is a mathematically-based symbolic-algebraic method that has eased the learning and teaching of the double-black node removal, coloring and balancing* of RB trees.

## 5   Conclusions

The deletion of a black node or the deletion of an internal red node and its subse-quent replacement by a black node in red-black (RB) trees causes a double-black (DB) node formation. In a RB tree, the DB node has no place. The challenges faced by students in the removal of the DB node and subsequent rebalancing of the tree with the allowed red and black node colors assignment to node is a fairly complex. After several years of teaching and research, this paper has addressed this difficulty in the teaching and learning of the DB node removal, and nodes recoloring by projecting a simplified and systematic algebraic-arithmetic proce-dure for both the removal of the DB node, and the recoloring of its siblings and adjacent (parent) nodes, subsequently. Our procedure has showed that there is a mathematically-based technique to ease the learning of DB removal in RB trees. This procedure which is a bottom-up approach that starts from the point of the deleted black node or the replacement node. This research work has stated six symbolic-algebraic formulas and demonstrated with illustrations the symbolic addition of a color and its inverse, $+B$ and $-B$; and the symbolic addition and inverse of the color red $+R$ and $-R$, respectively. This was then supported by an operational algorithm after testing with several RB tree structures. The illus-trations and testing showed that the symbolic-algebraic method conforms with, firstly; the conventional removal of DB node and node recoloring algorithm in RB tree visualization tools; and secondly, supported students' in-depth learning and understanding of DB removal. The next stage of this work is to continue the research on the use of the symbolic-arithmetic formulas on RB trees that involves node rotation with different cases, and to look at the computational time of symbolic-algebraic algorithm relative to the conventional algorithm for DB node removal and tree balancing.

**Acknowledgements.** We would like to acknowledge all past and present stu-dents of Data Structures & Algorithms at the Wenzhou-Kean University (WKU) for their collaboration, practice, and testing of this simplified symbolic-algebraic method for satisfiability of the properties of the red-black tree.

## References

[1] Besa, J., Eterovic, Y.: A concurrent red–black tree. Journal of Parallel and Distributed Computing **73**(4), 434–449 (2013), ISSN 0743-7315, https://doi.org/10.1016/j.jpdc.2012.12.010

[2] Bounif, L., Zegour, D.E.: A revisited representation of the red-black tree. In-ternational Journal of Computer Aided Engineering and Technology **16**(1), 95–118 (2022), https://doi.org/10.1504/IJCAET.2022.119541

[3] Fredriksson, E.: Reducing CPU scheduler latency in Linux. Bachelor the-     sis, Umeå University (2022), URL http://www.diva-portal.se/smash/get/diva2:1630380/FULLTEXT01.pdf

[4] Galles, D.: Red Black Tree Visualization (2011), URL https://www.cs.usfca.edu/~galles/visualization/RedBlack.html

[5] Germane, K., Might, M.: Deletion: The curse of the red-black tree. Journal of Functional Programming **24**(4), 423–433 (2014), https://doi.org/10.1017/S0956796814000227

[6] Ghiasi-Shirazi, K., Ghandi, T., Taghizadeh, A., Rahimi-Baigi, A.: Revisiting 2-3 Red-Black Trees with a Pedagogically Sound yet Efficient Deletion Algorithm: The Parity-Seeking Delete Algorithm (Jun 2022), https://doi.org/10.48550/ARXIV.2004.04344, URL https://arxiv.org/abs/2004.04344

[7] Goodrich, M.T., Tamassia, R., Goldwasser, M.H.: Data Structures and Algorithms in Java. John Wiley & Sons, 6 edn. (2014), ISBN 978-1-118-77133-4

[8] Hanke, S.: The Performance of Concurrent Red-Black Tree Algorithms. In: Vitter, J.S., Zaroliagis, C.D. (eds.) Algorithm Engineering, pp. 286–300, Springer Berlin Heidelberg, Berlin, Heidelberg (1999), ISBN 978-3-540-48318-2, https://doi.org/10.1007/3-540-48318-7_23

[9] Hasanzadeh, M., Alizadeh, B., Baroughi, F.: The cardinality constrained inverse center location problems on tree networks with edge length augmentation. Theoretical Computer Science **865**, 12–33 (2021), ISSN 0304-3975, https://doi.org/10.1016/j.tcs.2021.02.026

[10] Jeong, M., Lee, E.: A Swapping Red-black Tree for Wear-leveling of Non-volatile Memory. The Journal of the Institute of Internet, Broadcasting and Communication **19**(6), 139–144 (Dec 2019), https://doi.org/10.7236/JIIBC.2019.19.6.139

[11] Kahrs, S.: Red-black trees with types. Journal of Functional Programming **11**(4), 425–432 (2001), https://doi.org/10.1017/S0956796801004026

[12] King, J.: Combining Theory and Practice in Data Structures & Algorithms Course Projects: An Experience Report. In: Proceedings of the 52nd ACM Technical Symposium on Computer Science Education, p. 959–965, SIGCSE '21, Association for Computing Machinery, New York, NY, USA (2021), ISBN 9781450380621, https://doi.org/10.1145/3408877.3432476

[13] Li, J., Xu, Y., Guo, H.: Memory organization in a real-time database based on red-black tree structure. In: Fifth World Congress on Intelligent Control and Automation (IEEE Cat. No.04EX788), vol. 5, pp. 3971–3974 Vol.5 (2004), https://doi.org/10.1109/WCICA.2004.1342243

[14] Liang, Y.D.: Introduction to Java Programming, Brief Version, Global Edition. Pearson Education, 11 edn. (2018)

[15] Liew, C.W., Nguyen, H.: Using an Intelligent Tutoring System to Teach Red Black Trees. In: Proceedings of the 50th ACM Technical Symposium on Computer Science Education, p. 1280, SIGCSE '19, Association for Computing Machinery, New York, NY, USA (2019), ISBN 9781450358903, https://doi.org/10.1145/3287324.3293823

[16] Nipkow, T.: Teaching Algorithms and Data Structures with a Proof Assistant (Invited Talk). In: Proceedings of the 10th ACM SIGPLAN International Conference on Certified Programs and Proofs, p. 1–3, CPP 2021,

Association for Computing Machinery, New York, NY, USA (2021), ISBN 9781450382991, https://doi.org/10.1145/3437992.3439910

[17] Pranesh, Deshpande, S.L.: Tree-Based Approaches for Improving Energy Efficiency and Life Time of Wireless Sensor Networks (WSN): A Survey and Future Scope for Research. In: Ranganathan, G., Chen, J., Rocha, Á. (eds.) Inventive Communication and Computational Technologies, pp. 583–590, Springer Singapore, Singapore (2020), ISBN 978-981-15-0146-3, https://doi.org/10.1007/978-981-15-0146-3_55

[18] Qiaoyu, L., Jianwei, L., Yubin, X.: Performance Analysis of Data Organization of the Real-Time Memory Database Based on Red-Black Tree. In: 2010 International Conference on Computing, Control and Industrial Engineering, vol. 1, pp. 428–430 (2010), https://doi.org/10.1109/CCIE.2010.113

[19] Sanderson, C., Curtin, R.: A User-Friendly Hybrid Sparse Matrix Class in C++. In: Davenport, J.H., Kauers, M., Labahn, G., Urban, J. (eds.) Mathematical Software – ICMS 2018, pp. 422–430, Springer International Publishing, Cham (2018), ISBN 978-3-319-96418-8, https://doi.org/10.1007/978-3-319-96418-8_50

[20] Sedgewick, R.: Left-leaning Red-Black Trees. In: Dagstuhl Workshop on Data Structures, vol. 17 (Sep 2008), URL https://sedgewick.io/wp-content/themes/sedgewick/papers/2008LLRB.pdf

[21] Seidametova, Z.: Some methods for improving data structure teaching efficiency. Educational Dimension **58**, 164–175 (Jun 2022), https://doi.org/10.31812/educdim.4509

[22] Wu, D., Guo, P., Zhang, C., Hou, C., Wang, Q., Yang, Z.: Research and Practice of Data Structure Curriculum Reform Based on Outcome-Based Education and Chaoxing Platform. International Journal of Information and Education Technology **11**(8), 375–380 (2021), ISSN 2010-3689, https://doi.org/10.18178/ijiet.2021.11.8.1537

[23] Xhakaj, F., Liew, C.W.: A New Approach To Teaching Red Black Tree. In: Proceedings of the 2015 ACM Conference on Innovation and Technology in Computer Science Education, p. 278–283, ITiCSE '15, Association for Computing Machinery, New York, NY, USA (2015), ISBN 9781450334402, https://doi.org/10.1145/2729094.2742624

[24] Zegour, D.E.: Improving the Red-Black tree delete algorithm (Jul 2022), https://doi.org/10.21203/rs.3.rs-1194654/v3

[25] Zhang, H., Liang, Q.: Red-Black Tree Used for Arranging Virtual Memory Area of Linux. In: 2010 International Conference on Management and Service Science, pp. 1–3 (2010), https://doi.org/10.1109/ICMSS.2010.5575666